\documentclass[11pt]{article}
\usepackage{amsmath,amssymb,color,epsfig,cite}
\usepackage{graphicx}
\usepackage{setspace}
\usepackage{mathrsfs}

\usepackage[english]{babel}
\usepackage{inputenc}
\usepackage{amsfonts}
\usepackage{soul} 
\usepackage{hyperref}
\usepackage{subfigure}
\usepackage[normalem]{ulem}
\usepackage{array}
\usepackage{authblk}

\newcommand{\hoch}[1]{$\, ^{#1}$}

\begin{document}

\vspace{15pt}

{\Large {\bf A Geometric Framework for Interstellar Discourse on Fundamental Physical Structures}}
\vspace{15pt}

{\bf Giampiero Esposito \hoch{1,2,}, Valeria Fionda\hoch{3}}
\vspace{10pt}

\hoch{1}{\it Dipartimento di Fisica ``Ettore Pancini'',
Complesso Universitario di Monte S. Angelo, Via Cintia Edificio 6, 80126 Napoli, Italy}

\hoch{2}{\it INFN Sezione di Napoli, 
Complesso Universitario di Monte S. Angelo, Via Cintia Edificio 6, 80126 Napoli, Italy}

\hoch{3}{\it Department of Mathematics and Computer Science, University of Calabria, Italy}

\vspace{20pt}

\underline{ABSTRACT}
This paper considers the possibility that abstract thinking and 
advanced synthesis skills might encourage
extraterrestrial civilizations to accept communication with
mankind on Earth. For this purpose, a notation not relying upon 
the use of alphabet and numbers 
is proposed, in order to denote just
some basic geometric structures of current physical theories: vector fields,
$1$-form fields, and tensor fields of arbitrary order. An advanced
civilization might appreciate the way here proposed to achieve a
concise description of electromagnetism and general relativity, 
and hence it might accept the challenge of responding to our signals.
The abstract symbols introduced in this paper to describe the basic structures 
of physical theories are encoded into black and white bitmap images that 
can be easily converted into short bit sequences and modulated on 
a carrier wave for radio transmission.

\noindent
\thispagestyle{empty}

\vfill

gesposito@na.infn.it, valeria.fionda@unical.it

\pagebreak

\section{Introduction}

The modern theories of fundamental interactions have developed
powerful tools for addressing some outstanding problems, i.e., what
is the ultimate nature of gravity, and how to achieve unification
of all fundamental interactions and guiding principles of physics.
At a deeper level, however, 
yet another problem arises, i.e., how to develop
a language which is universally understood by the scientific 
community. This is not a purely formal issue. For example, since 
the language of physics is mathematics, we might consider the
question: if we use alphabets in order to express physical laws in
mathematical form, what tells us that $X$ is a vector field, rather
than a topological space, or a point of a metric space, or a
$1$-form field? At present, several authors find a way out by using
the so-called abstract indices, e.g., $X^{a}$ for vector fields,
but this does not truly solve the problem, because it needs a second
family of symbols taken from the alphabet, and on our planet no unique
alphabet exists. Moreover, upstairs Latin indices are also used to 
denote families of $1$-form fields, e.g., the tetrads
$$
e^{a}=\sum_{\mu} e_{\; \mu}^{a} dx^{\mu},
$$
nor can we distinguish between a vector and a vector field unless we
introduce for the former a letter to denote the point $p$ at which 
the vector is considered. Hence one writes $X_{p}$ for the vector, and
$X$ for the vector field, which consists of a smooth (of class
$C^{\infty}$ or less) assignment of $X_{p}$ as $p$ is varying on
the manifold of interest. Thus, unless we use twice the alphabet, no
clear understanding of our notation is available, even though we
refrain us completely from using coordinate descriptions. 

At this stage, this seems to be just a well motivated but minor issue
for modern science. However, if we consider the open problem of
CETI, i.e., communication with extra-terrestrial intelligence, the
associated perspective might change. So far, mankind has sent
signals that contain a synthesis of our current understanding of
biology, chemistry and physics. 
In our opinion, however, the following question is also important: 
if an advanced extraterrestrial
civilization receives radio transmissions, what 
features would convince them that it is worth spending time
in trying to decode them? 
In our opinion, the more advanced they are, the more they would prefer
to detect clear indication of original thinking in outlining the
basic structures of the concepts we use, rather than the detailed 
operations that we perform upon them. Although the distances involved
are so large that it might take thousands of years before such an 
event takes place, its relevance in a cosmic perspective cannot be
easily denied (cf. the literature on the Fermi paradox
\cite{Gray,Brin,Webb}).

Our proposal is an abstract and simple notation, here called Physics Lingo, 
that plays the role of a bridge, connecting the abstract language of physics with 
the universal power of visual imagery. Rather than relying uniquely on traditional 
mathematical symbols and equations, Physics Lingo harnesses the expressive 
capacity of images by offering a unique perspective on transforming physical 
concept and theories into tangible visual representations. At the core of Physics 
Lingo lies the concept of encoding physics principles and theories, represented 
by abstract symbols, into digital bitmap images. By putting the emphasis
upon visualization, we are assuming that the receivers of the message
have some analog of human visual perception (at the risk of limiting
the kinds of extraterrestrial civilizations that would understand the message). 
This representation of abstract symbols into digital bitmap images 
is not new, but its application to the notation that we 
describe in Sec. 2 is the innovative contribution of our paper.
Each bitmap image conveys specific 
meanings and forms the foundation for a visual language that can overcome language 
barriers and communicate the richness of physics across the cosmos. The beauty of 
Physics Lingo lies not only in its ability to encode complex physics concepts, but 
also in its potential to serve as a method of extra-terrestrial interstellar communication. 
As we seek to establish contact with intelligent beings from distant star systems, 
the simplicity and universality of the bitmap images offer a promising platform for communication.

Section $2$ describes the special notation without alphabet that we propose, 
while some key aspects of interstellar communication are discussed in Sec. 
$3$, and Sec. $4$ studies the transmission technique. Concluding remarks are presented in Sec. $5$.

\section{Physics Lingo: A Language for Cosmic Dialogue}

The first non-trivial problem is: which branch of mathematics
do we want to describe in the original synthesis of the structures
and concepts that we use in our research and our teaching?
As far as we can see, the basic geometric objects provide a good
candidate in this respect. Our prescriptions are as follows.
\vskip 0.3cm
\noindent
(i) Tangent vectors: In differential geometry, tangent vectors are
first defined as directional derivatives of a function along a
curve. Eventually, one finds that a tangent
vector $X_p$ is an equivalence class of curves tangent at a point $p$.
An assignment of class $C^{\infty}$ (i.e., continuous together with
the derivatives of any order) of tangent vector as the point $p$ 
varies on the manifold $M$ is said to be a vector field.
In this section, at the risk of slight repetitions of our remarks in
the Introduction, we point out what follows.

Modern physical theories have a geometric nature, and a very important
concept in global differential geometry is the one of a vector field
given above, out of which one obtains $1$-form fields and all tensor
fields. But all notations conceived by mankind suffer from drawbacks.
The letters $X$ and $Y$ do not mean, by themselves, that we are considering
vector fields, unless we specify it in words, nor is $T$ enough to
denote a tensor field of type $(r,s)$. On the other hand, if we use
index notation, we run the risk of confusing a tensor field with its
components in a given coordinate system, which is a serious mistake.
However, an advanced civilization must have discovered the concepts
of contravariance and covariance, and the use of arrows (see below)
as a substitute of index-free or index notation may be understood
if a simple message is sent. Furthermore, we propose a simple notation 
based on round brackets (see below) in order to replace our
alphabet(s) or any universal alphabet that one might want to use.
Thus, rather than writing $X_{p}$ for the tangent
vector at $p \in M$, which is an element of the tangent space 
$T_{p}(M)$, we write $(\; \; \;)_{\odot}^{\uparrow}$ where the empty
space within square brackets is a substitute for $X$, the dot
downstairs is a substitute for the point $p$, and the arrow 
upstairs distinguishes a vector from its dual concept of $1$-form
or covector. We are therefore introducing a substitution map
\begin{equation}
X_{p} \longrightarrow \; (\; \; \;)_{\odot}^{\uparrow}
\label{(1)}
\end{equation}
Note that, with our notation, round brackets do not occur alone,
but are accompanied by arrows and dots. Thus, the receiver might
be led to assume that brackets are a substitute for whatever
letter or symbol one might want to use. Indeed, an eminent
author \cite{Penrose} has developed long ago an advanced notation
which replaces tensors and operations on tensors by graphical
symbols, but the associated diagrams would require far more pixels
and are therefore less suitable, in our opinion, for the purpose
of making it easier for the receiver to decode the message. 

Tensor calculations are not simplified by our
notation, but students understand it
without difficulties, based on the teaching
experience of one of us. Our language is promising but
still incomplete. For example, the equality symbol is missing,
but on the other hand 
this should not be too surprising, since such a concept does not
have an universal meaning: a set theorist might say that the
abstract sets $A$ and $B$ are equal if their difference is the
empty set, whereas an expert in measure theory might want to
identify $A$ and $B$ if the measure of their difference vanishes.

The concept of symmetric tensor of type $(0,2)$ might be expressed
by writing first a round bracket accompanied by two arrows of
unequal length pointing downwards, and then repeating the same 
diagram but with the roles of the two arrows being interchanged.
Such a repetition might suggest that, for the sender, the two
diagrams represent equivalent concepts. Such a communication
reduced to the minimal set of technical ingredients might
stimulate intellectual curiosity of the receiver.

Our arrows might be replaced by geometric figures placed
upstairs or downstairs (e.g., a star or a quadrangle), but
the resulting digital transmission of signals would not
get simplified. Thus, our notation is not necessarily
anthropocentric, as far as we can see.

If it were easy to correspond in space, we could say that
our framework would only appear as 
a candidate for what should be part of a longer message,
initially devoted to more basic concepts and informations
about the Earth and life, and to the way in which
mankind built the current knowledge of Nature. 
But the length and time scales
involved, and the difficult process of decodification, 
might encourage a non-standard choice of priorities
in what we try to transmit. To the extent that the receiver
is really very advanced, he/she might enjoy a message
that displays an original synthesis and relies upon
abstract thinking. As far as we know, this is a novel idea.
We have focused on geometry because it offers appropriate
tools and concepts, but of course other branches of
mathematics offer opportunities which deserve a
thorough consideration.
\vskip 0.3cm
\noindent 
(ii) Vector fields: in this case, our substitution map acts
according to
\begin{equation}
X \longrightarrow \; (\; \; \;)^{\uparrow}
\label{(2)}
\end{equation}
\vskip 0.3cm
\noindent
(iii) $1$-forms: we require that
\begin{equation}
\omega_{p} \longrightarrow \; (\; \; \;)_{\odot}^{ \downarrow}
\label{(3)}
\end{equation}
where the arrow downstairs denotes the concept dual to the one
in item (i).
\vskip 0.3cm
\noindent
(iv) $1$-form fields: we write
\begin{equation}
\omega \longrightarrow \; (\; \; \;)_{\downarrow}
\label{(4)}
\end{equation}
where the dot has disappeared, as is the case in Eq. (2),
since we are dealing with a smooth assignment of $1$-form as
the point $p$ is varying on the manifold $M$. 
\vskip 0.3cm
\noindent
(v) Tensor fields: a tensor field of type $(2,3)$ takes the form
\begin{equation}
(\; \; \;)_{\; \; \; \downarrow \downarrow \downarrow}^{\uparrow \uparrow}
\label{(5)}
\end{equation}
and, in particular, the Riemann curvature tensor, which is
of type $(1,3)$, can be denoted by
\begin{equation}
(\; \; \;)_{\; \downarrow \downarrow \downarrow}^{\uparrow}
\label{(6)}
\end{equation}
\vskip 0.3cm
\noindent
(vi) The spacetime manifold $(M,g)$, one of the great achievements
of Einstein, can be expressed in the form
\begin{equation}
\Bigr((\; \; \; \;)(\; \; \;)_{\downarrow \downarrow}\Bigr).
\label{(7)}
\end{equation}
(vii) Vacuum Maxwell theory in curved spacetime can be described
by the potential $1$-form field $A$, its gauge curvature $2$-form
field $F=dA$, and the Hertz \cite{Bouas,Cohen} 
$2$-form $P$ such that

$$\delta P=*d*P=A.$$

The simple but non-trivial triplet
\begin{equation}
(P,A=\delta P,F=dA)
\label{(8)}
\end{equation}
displays the electromagnetic potential, a $1$-form field $A$,
which lies in between the $2$-form field $P$ and the $2$-form
field $F$. The coderivative of $P$ yields $A$, while the exterior
derivative of $A$ yields $F$. Now, although there is no obvious
graphical representation for coderivative and exterior derivative,
we can nevertheless say that the underlying mathematical 
structure is the triplet below:
\begin{equation}
(\; \; \;)_{\downarrow \downarrow} \; \; \; \; \; \;
(\; \;)_{\downarrow} \; \; \; \; \; \; 
(\; \; \;)_{\downarrow \downarrow}
\label{(9)}
\end{equation}
We assume that, for the
educated receiver relevant for the CETI program, this might be
enough for him/her to understand that this is a shorthand notation
for the content of Eq. (8). In order to achieve this, the
logical order in which our diagrams appear (at this stage, the reader 
may have a first look at our Figure $2$) is of
crucial importance: first, vectors at a point with the associated
vector fields, and the counterpart consisting of $1$-forms at
a point with the associated $1$-form fields. Next, tensor fields
of arbitrary order, then the unification of space and time by
means of the spacetime manifold, and eventually electromagnetism
and general relativity. Such a logical order might help the
receiver in understanding the global meaning and 
appreciating our synthesis skills.
\vskip 0.3cm
\noindent
(viii) Tensor affinities are unfortunately denoted in the literature 
in the same way
as tensor components, even though they do not transform in a
homogeneous way under diffeomorphisms. Within our concise language,
we propose instead to denote them in the form
\begin{equation}
(\; \; \;)_{\; \; \; \sim \sim}^{\sim}
\label{(10)}
\end{equation}
The educated reader might understand that the resulting objects
resemble tensor components, but do not transform tensorially 
under diffeomorphisms. {\it Only arrows are appropriate for 
visual description of tensorial behaviour}. We stress, however,
that Eq. (10) is written only for completeness, but we do not
propose to include it in a first message for interstellar
communications. The symbol $\sim$, when used in a context where
the vast majority of symbols accompanying round brackets are just
arrows pointing upwards or downwards, will manifestly suggest that
it represents objects of different nature but similar graphical form.

Of course, our notation does not tell the receiver how we perform operations
by using such geometric structures, but it is an indication
(among the many conceivable) of original thinking that might attract
the interest of the receiver. To the extent that the receiver is
truly advanced, the indication of synthesis skills and original
thought might play a role in motivating the effort to understand
our message.

\subsection{Comparison with other methods of ETI communication}

The work in Ref. \cite{Jiang} has developed a novel binary-coded
message for transmission to extraterrestrial intelligences in the
Milky Way galaxy. Such a message includes basic mathematical and
physical concepts to establish a universal method of communication
followed by information on the biochemical composition of life on
Earth, the Solar System's time-stamped position in the Milky Way
relative to known globular clusters, as well as digitized descriptions
of the Solar System, and Earth's surface. Eventually, the message 
displays digitized images of the human form, along with an explicit
invitation for any receiving intelligences to respond. This work
relies upon the valuable language created in Ref. \cite{Dutil}.

The difference between our work and these outstanding references
is twofold. On the one hand, we do not focus on summarizing binary
and decimal systems, mathematical operations, particle physics, 
spectrum of hydrogen atom, DNA structures, illustrations of the 
human form, Solar System and Earth illustration, composition and
characteristics of Earth. On the other hand, we avoid using alphabets
of any sort, because we believe that the following considerations
can be made:
\vskip 0.3cm
\noindent
(i) One cannot be sure that an advanced civilization might be willing
to spend time in learning a new universal alphabet.
\vskip 0.3cm
\noindent
(ii) The use of graphical symbols is more direct. The more advanced
the extraterrestrial intelligence, the more they are likely to
understand that brackets enclosing empty spaces, and supplemented
by arrows pointing upwards or downwards (the occurrence of such
arrows being of crucial importance) are a substitute 
for using letters (but we are aware that we cannot prove to be right) 
which have however different shape in their planet and in our planet. 

Our approach is therefore, by construction, complementary to the
outstanding work in Refs. \cite{Jiang,Dutil}. Since it might take
thousands of years for a message to reach even the first of its
possible destinations \cite{Jiang}, it is not easy to decide whether
or not our scheme is going to be successful.

Since many valuable works have been published on various aspects
of interstellar communication, we bring here a list to the attention
of the general reader, as follows. A language based on the fundamental
facts of science \cite{Devito}, symbol systems and pictorial
representation \cite{Diederich}, yet other efforts to build a
language and select the contents of our messages 
\cite{Freudenthal,Heidmann,Lemarchand,Musso,Nieman,Sagan78,Sagan85,Vakoch}. 

\section{Some peculiar features of interstellar communication}

When talking about interstellar communication, the general reader
should bear in mind that, apart from the Pioneer plaques and Voyager records, 
every interstellar message has used microwave frequencies as the
communication medium. Attention is restricted to a relatively
narrow band, ranging from about $1$ GHz to $10$ GHz.
This is said to be the microwave window, because it is one of the
few regions of the electromagnetic spectrum that is unimpeded
by Earth's atmospheric gases \cite{Oberhaus}.
The radio signals that we send have to travel through
energy-absorbing clouds of interstellar dust and gas, hence 
they will be quite faint by the time they arrive at their
destination.
The probability that a message from Earth is detected
is very low, because of the large number of possible
frequencies to broadcast on.  
An alternative to microwave signals consists of optical communication
based upon laser transmission, which however has a far higher energy requirement.
Moreover, optical telescopes also require sophisticated
equipment adaptations in order to make them capable of transmitting
laser signals. Other important issues are the choice of wavelengths at which
we should search for signals from extraterrestrial intelligence \cite{PNA1},
and the estimate of mean number of radio signals crossing the Earth \cite{PNA2}.

\begin{figure}[!h]
    \centering
    \includegraphics[width=11cm]{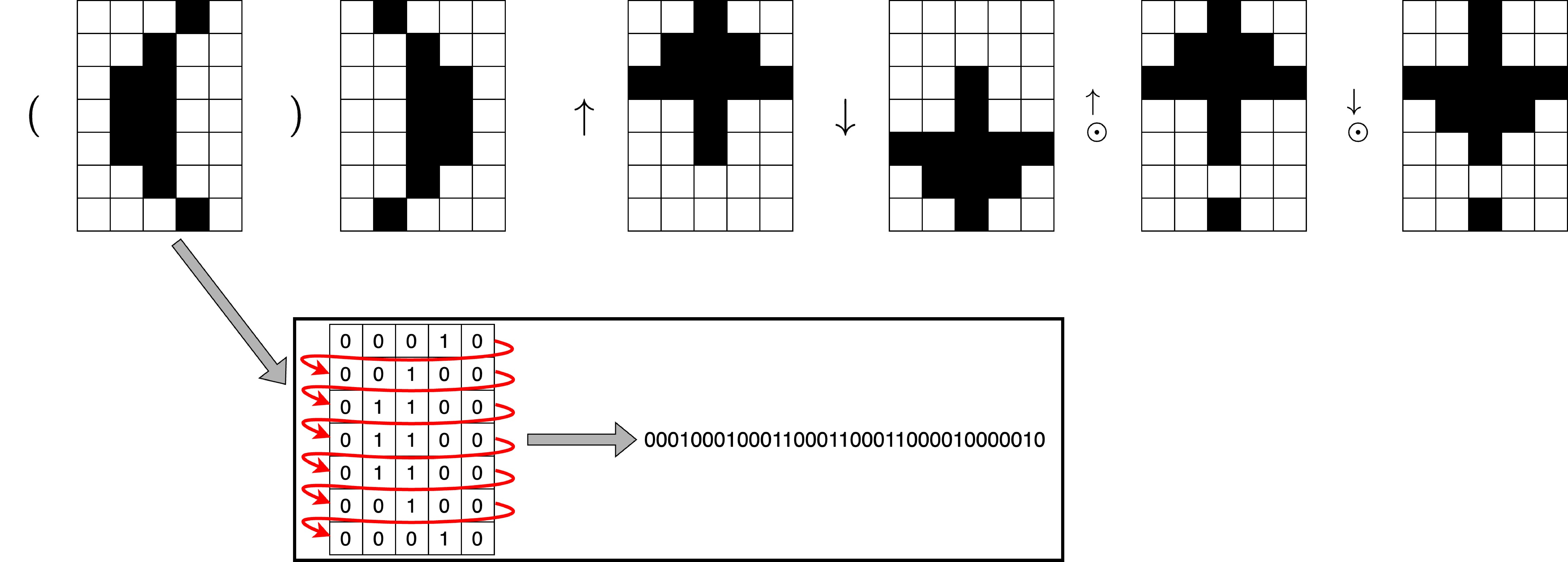}
    \caption{Bitmap representation of the symbols in Eqs. (1)-(7) and (9) 
and bit sequence encoding for the symbol '$($'. }
    \label{fig:bitmap1}
\end{figure}

\section{Encoding and Transmitting Physics Words}

This section focuses on realizing digital signals that
enable the receiver to build the simple images
described by Eqs. [1]-[7] and [9].
For this purpose, we rely upon the bitmap representation, that has gained significant 
attention by virtue of its simplicity and efficiency in data storage and transmission. 
Several studies focused on such a representation of digital images, by underlining 
its applicability to different fields (e.g., medicine~\cite{Deserno}, 
astronomy~\cite{Dimov}, document analysis~\cite{Rivest}) and 
providing useful instruments to improve representation quality~\cite{Sarkar} or 
processing capabilities~\cite{Guibas}. 

The bitmap representation of a black and white image subdivides the image in a 
fixed number of points (i.e., into a grid of small squares called pixels) and 
associates to each point or pixel a value that is either black or white. Each pixel 
is then assigned a binary value, where 1 represents black and 0 represents white. 
Figure~\ref{fig:bitmap1} shows how the symbols reported in Eqs. [1]-[7] and [9] 
can be represented by grids that are 5 pixels wide and 7 pixels tall, hence the bitmap 
representation for each symbol is a 7x5 grid. The pixels stored in the bitmap 
grid can be then converted into a sequence of bits, as depicted in Figure~\ref{fig:bitmap1}, 
by concatenating the bits of each row from the top to the bottom of the bitmap grid. 
Each bit in the sequence corresponds to a pixel in the image. To ensure the 
receiver accurately understands the dimensions of the grid, a brief pause will be 
inserted between the transmission of each row's sequence of bits, delineating the 
rows and aiding in reconstructing the original 7x5 grid accurately. 
Our selection of a 5x7 grid was initially guided by considerations of bitmap font design, 
where such small grids are a common standard due to their efficiency in balancing detail 
and simplicity for character representation. Such a size allows for sufficient complexity 
to represent a range of symbols while maintaining minimalism to avoid unnecessary confusion.
In interstellar communication there is the added difficulty of creating a 
message for extraterrestrial intelligence without an established 
communication protocol and shared language or context. 
In this respect some standard strategies could be employed 
to indicate that a message contains a 
sequence of bitmaps meant to be interpreted as images: 
\textit{(i)} a sequence of bitmaps could be arranged in a grid 
with dimensions that are prime numbers, suggesting intentional structuring that an 
intelligent recipient might recognize as non-random; \textit{(ii)} repeating each 
small bitmap or a sequence of bitmaps multiple times can signal that 
the pattern is significant and intended as a communication message. Note that, 
our $5 \times 7$ grid representation of the binary sequences provides a 
unique factorization akin to the approach used in the Arecibo message, as $35$ is the product 
of two prime numbers, $5$ and $7$, and there are no other factors to consider. 

\begin{figure}[!h]
    \centering
    \includegraphics[width=11cm]{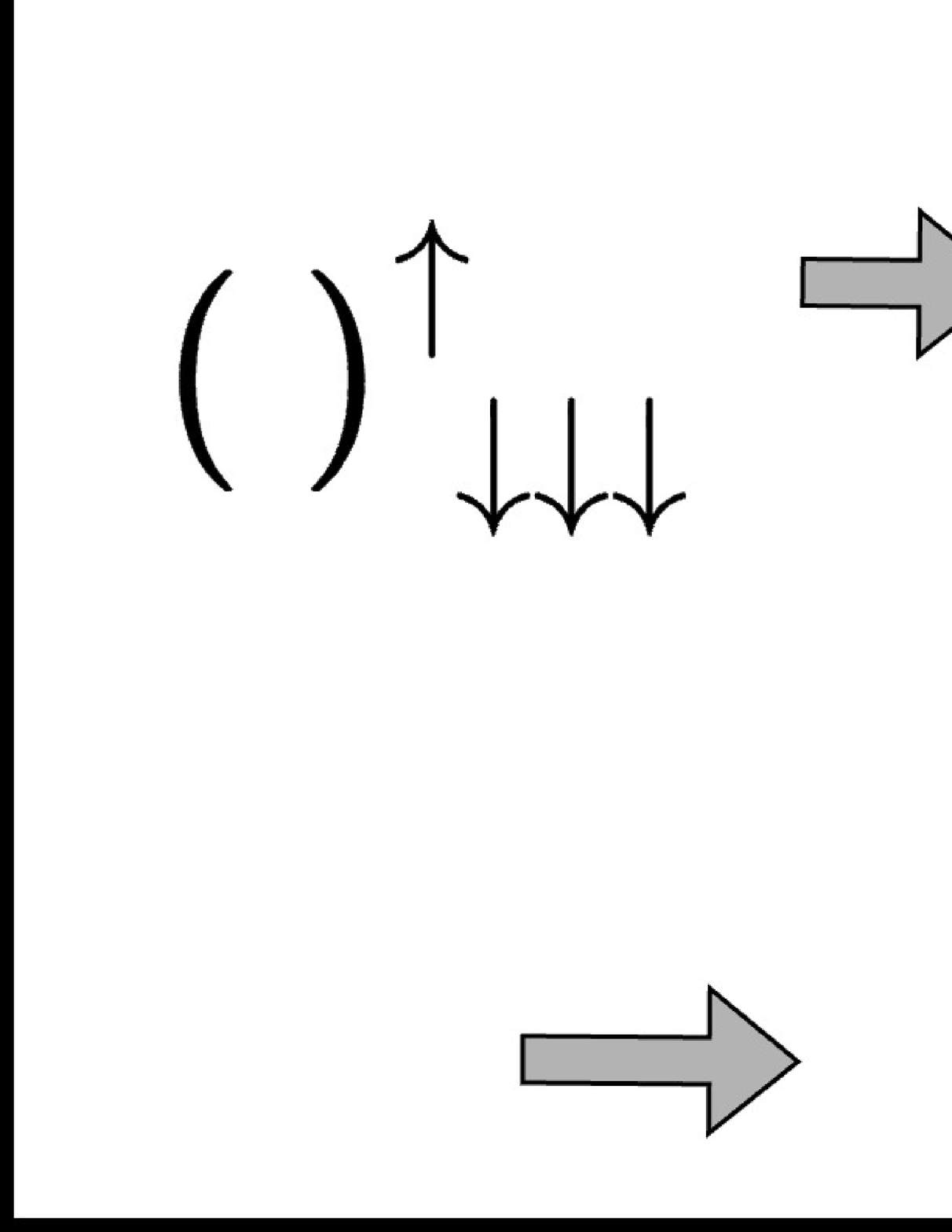}
    \caption{Bit encoding for the images in Eqs. (6), (7) and (9). The first five 
rectangles represent some basic geometric concepts, i.e. a vector, a vector field, 
a $1$-form, a $1$-form field, and a tensor field of type $(2,3)$, respectively. 
    The sixth rectangle describes the spacetime manifold, while the seventh rectangle is 
devoted to the electromagnetic theory of light, and the eighth rectangle focuses 
on the Riemann curvature tensor.}
    \label{fig:bitmap3}
\end{figure}

\subsection{Signal modulation}

In order to transmit these data as a radio signal, the binary sequence needs 
to be modulated onto a carrier wave. Indeed, binary modulation enables the efficient 
transmission of digital information over long distances. The process begins with 
the binary signal (the sequence of 1s and 0s representing the digital information) 
that is modulated onto a radio wave carrier by using existing modulation schemes such 
as amplitude-shift keying (ASK), frequency-shift keying (FSK), or phase-shift keying 
(PSK). In ASK, the carrier's amplitude is varied to represent the binary data, with 
one amplitude denoting a '1' and another amplitude representing a '0'. FSK modulates 
the carrier's frequency, assigning different frequencies to '1' and '0'. PSK, 
on the other hand, alters the phase of the carrier wave to encode the binary signal. 
With the help of these modulation techniques, the binary signal is combined with the 
radio wave carrier, allowing it to propagate through the air.

In order to generate the signals corresponding to sequences of symbols such as 
the ones reported in Eqs. (1)-(7) and (9), we concatenate in a single sequence the bit 
sequences corresponding to each symbol. Figure~\ref{fig:bitmap3} summarizes in the 
last three rectangles the bit encoding for the bitmap images in 
Eqs. (6), (7) and (9).

\subsection{Generality of the approach}

The presented technique, which involves converting black and white images into a 
bitmap representation and then encoding them into bit sequences that are modulated 
onto a carrier wave, exhibits an encouraging generality in its applicability. In fact, 
it can be effectively employed for transmission of any black and white image, 
if the appropriate size of the grid is selected. By adjusting the grid size, 
the technique can accommodate images of different complexity by handling 
different levels of detail. 
As a final note, we would like to stress that when symbols are encoded into
bitmap images, it is advantageous to utilize a lower-dimensional grid;
indeed, this makes it possible to save resources required for transmission. By
minimizing the pixel count in bitmap images, the amount of data to be
transmitted is reduced, leading in turn to a more resource-efficient
communication process. This is especially critical in space
communications, where bandwidth and power are precious commodities.

\section{Concluding remarks}

The topic of interstellar communications is still receiving a careful consideration
in the modern literature, from different perspectives
\cite{Seth2010,Seth2011,Hippke2018,Hippke2020,Berera2020,Hippke2021}.
In particular, we agree with the author of Ref. \cite{Hippke2020} that other
technological life exists; a relevant fraction of ETI wishes to explore the
universe; travel velocities are sufficiently high to allow for probes that
arrive at their destination intact, with a useful lifetime remaining; last,
but not least, a relevant fraction of ETI appreciate data of some kind.
Moreover, we assume that the receiver of messages from Earth is probably
thinking nature in abstract geometrical language.
Within this framework, our original contribution is, as far as we can see, twofold:
\vskip 0.3cm
\noindent
(i) We have first proposed a concise notation for describing some basic geometric 
concepts used in modern physical theories \cite{Esposito2023}.
\vskip 0.3cm
\noindent
(ii) We have then shown in detail how to obtain bit encoding for the images in 
Eqs. (1)-(7) and (9), which might be exploited for interstellar communication, since the resulting
synthesis skills might attract the attention of advanced civilizations which receive
signals sent from the Earth. We hope 
that the unification of space and time as depicted in the sixth rectangle of Fig. 2, the
electromagnetic theory depicted in the seventh rectangle therein, and the Riemann curvature 
represented in the eighth rectangle, will encourage careful consideration
on the side of the receivers.

By working along similar lines, other original concise notations for scientific concepts
might be hopefully developed, and a new longterm research program might be conceived.
In particular, it might be challenging to perform a visualizable synthesis of
mathematical and physical knowledge achieved so far. Such a synthesis might be
the result of dedicated efforts by several scientists, and hence
cannot be obtained in the present paper, whose aim has been instead to build a geometric 
framework for interstellar communication. We can only express the hope that, 
if ever reached, a visualizable synthesis of human knowledge would 
make it possible to choose examples for interstellar
quantum communications \cite{Hippke2021} as well.
Although a more advanced civilization might have reached the stage where they 
depend substantially on artificial intelligence \cite{Seth2010},  we hope that 
the original synthesis initiated in this paper will enhance future efforts 
in achieving interstellar communication.
The science of extraterrestrial language is a rapidly growing field,
for which we refer the reader to the work in Ref. \cite{VP2024}.  
The distances and time scales involved make it hard to select the appropriate path.

\end{document}